# Reproductive trends, habitat type and body characteristics in velvet worms Onychophora)

Julián Monge-Nájera
Centro de Investigación General, (UNED. *Mailing address:* Biología Tropical. Universidad de Costa Rica. San José. Costa Rica

**Abstract:** A quantitative analysis of several onychophoran characteristics shows that in habitats with lower rain levels females reproduce at an older age, are more fecund and tend to have reproductive diapause where rain does not exceed a mean of 200 cm/year. These habitat characteristics are associated with the southern family Peripatopsidae. Sex ratio and parental investment per young are not correlated with general environmental conditions. A comparison of 72 species showed that larger species are often more variable in morphometry, but species with the longest females do not always have the longest males. Larger *Peripatus acacioi* females (Peripatidae: Brazil) produce more and heavier off spring. Intrapopulation morphology was studied in 12 peripatid species for which samples of between II and 798 individuals were available. In general, within populations the females are more variable than males in' length and weight, but similarly variable in the number of legs. The number of legs has a low variability (1.73-2.45%). length is intermediate (22.4-25.3%) and weight is very variable (49.41-75.17%). When sexes are compared within a population, females can have 14-8.9 % more leg pairs, and be 47-63 % heavier and 26 % longer than males.

The analysis of reproductive trends in the phylum Onychophora has been limited by the fragmentary nature of the data (Read 1985).The general conclusions so far have been that most studied species make a relatively high parental investment in their offspring (particularly those from the Neotropics) and that reproductive diapause does not seem to follow a simple geographical pattern (Read 1985, Ruhberg 1985. Morera *et al.* 1988).

What ecological factors may have influenced such characters as age at first reproduction, parental investment and fecundity have received little if any attention (Read 1985, Morera *et al.* 1988, Havel *et al.* 1989). Similarly, any possible associations between reproduction and body characteristics have only been considered in a qualitative way *(e.g.* Ghiselin 1974). There is no published regression of maternal mass, fecundity and investment in a large onychophoran sample, but the data have been available for years in the report of Lavallard and Campiglia (1975) and will be re-analyzed here.

The Peripatidae, one of the two families of the phylum, has a tropical distribution and in comparison with the temperate Peripatopsidae. has species of a larger body size and more legs which vary significantly in number (Bouvier 1905, Ruhberg 1985).

There are few studies analysing populations rather than species'ranges (reviews in Read 1985 and Ruhberg 1985). Nevertheless, it is known that within a species, females tend to be larger *(e.g.* 2.7 times heavier) and may be 50 % longer than males (Campiglia and Lavallard 1973, Monge-Nájera and Morera 1994). The growth rate varies with the species and is higher in females and in captivity *(e.g. Peripatus acacioi* 0.185-0.676, *Epiperipatus edwardsii* 0.65. *Epiperipatus imthurni* 4.77, *Macroperipatus torquatus:* females 5.7, males 3, general in the wild 1.62, all units are mg/day: Read 1985).

Although the original proposal of using the number of legs to distinguish species (De Blain ville, in Gervais 1838) was unjustified (Bouvier 1905), the character is often useful to distinguish the sexes within a species (Bouvier 1905, Lavallard and Campiglia 1973). Males have a lower number of legs (Gaffron 1885, Lavallard and Campiglia 1973, Monge-Nájera and Morera 1994) and show more cases of asymmetry, in which one side of the body has at least one additional leg (there is no right or left trend: Lavallard and Campiglia 1973).

A weak statistical test failed to discover a relationship between weight and the number of legs in *Peripatus acacioi* (Campiglia and Lavallard 1973), and the same resulted from stronger tests applied to *Macroperipatus torquatus* and *E. trinidadensis* (Read 1985). In contrast, heavier newborn *E. imthurni* have more legs (Read 1985). The reason for this variability is unknown but suggests different selective pressures upon species.

This paper, based on simple statistical tests, examines the relationship between habitat type and reproductive and morphological characteristics of several onychophoran species, considering species characteristics (both families) and variations within populations (Peripatidae).

MATERIAL AND METHODS

Analysis at the species level: Information from the literature was compiled (Tables 1-3) and analysed with (a) Spearman rank correlations, whose coefficients are indicated by a "c" in the text (ordinal and continuous variables: "number of legs", "body length", "age at first parturition", "fecundity" and "parental investment"). Only significant (p.<O.O5) correlations of 0.70 or higher were included (b) the Mann- Whitney U or Student's t tests (continuous and ordinal variables compared for the binomial variable "reproductive frequency"), or (c) Chi-square or Fisher Exact tests as required by sample size (cases involving the other variables, which are categorical). Data for categorical variables were pooled in two by two tables as fit for sample size. Males and females were analysed independently in each family, but data appear pooled when there was no statistical difference.

The data for length and number of leg pairs (Table 3) were taken from taxonomic descriptions and often refer to preserved specimens.

**Analysis within populations:** Morphometric data from around the world were taken from the literature and analysed with nonparametric tests, which are fit for this type of data (sources in Table 5).

RESULTS

**Family comparison:** When compared with the tropical family Peripatidae (Table 1), the southern Peripatopsidae show a higher frequency of reproductive diapause, longer female maturation (mean 30 months against 15 of Peripatidae) and higher fecundity (mean 23 young per year against 9 of Peripatidae). The age sample is too small for any statistical test, but fecundity differs significantly (Student's p<0.05). Parental investment per young is un significantly higher in Peripatidae (mean 6.9 % against 4.6% in Peripatopsidae, Student's I p>0.05).

Peripatid species have higher mean values for minimum length and for minimum and maximum number of legs in both sexes, and Peripatid females are more variable in number of legs than peripatopsid females (Table 4, Mann-Whitney U tests, p=0.0000-0.042). Peripatopsid males are longer (maximum length) and both sexes are more variable in length (Table 4, Mann-Whitney U tests, p=0.0000-0.03). There is no difference between families in maximum female length and in variability in the number of legs per male (Table 4, Man-Whitney U tests, p>0.05).

Character associations with habitat: In most cases, reproductive characters were not correlated among themselves or with environmental conditions (Table 1). The exceptions were (1) a tendency of females to begin reproducing at a younger age in moister habitats (Spearman p<O.O5) and (2) the preponderance of reproductive diapause when mean rainfall is below 200 cm per year (Fisher p=O.0001). The diapause was also associated with two climate types: "Warm with Dry Winter" and "Damp Temperate" (Fisher p=0.O34). Parental investment and sex ratio do not seem to be correlated with habitat or family (Tables 1 and 2).

TABLE 1
Reproductive characteristics and environmental conditions for 24 onychophoran species

| Taxa and characters | PI | RF | AF | FEC | QV | VG | PH | CL | TP | MR | RA |
|---|---|---|---|---|---|---|---|---|---|---|---|
| Peripatidae | | | | | | | | | | | |
| Epiperipatus hilkae | 5.95 | 2 | — | 4< | 1 | 1 | 1 | 1 | 26.6< | 4 | 5 |
| Plicatoperipatus jamaicensis | 12 | 2 | — | 10-20 | 1 | 1 | 1 | 2 | 26.6< | 3 | 5 |
| Peripatus acacioli | 7.95 | 1 | 15-23 | 1-8 | 6 | 4 | 1 | 3 | 21.1 | 1 | 4 |
| Epiperipatus brasiliensis | 6.5 | 2 | — | — | 1 | 1 | 1 | 2 | 26.6< | 4 | 5 |
| Epiperipatus isthmicola | 6.7 | 2 | — | — | 1 | 1 | 1 | 2 | 26.6< | 4 | 5 |
| Macroperipatus torquatus | 5.3 | 2 | 20.5 | 7.4 | 1 | 1 | 1 | 2 | 26.6< | 3 | 5 |
| Epiperipatus imthurni | 5.1 | 2 | 12.2 | 16.9 | 1 | 1 | 1 | 2 | 26.6< | 3 | 5 |
| Peripatus juliformis | 7.5 | 2 | — | — | 1 | 1 | 1 | 2 | 26.6< | 3 | 5 |
| Epiperipatus trinidadensis | 7.6 | 2 | — | — | 1 | 1 | 1 | 2 | 26.6< | 3 | 5 |
| Epiperipatus edwardsii | 4.4 | 2 | — | — | 1 | 1 | 1 | 2 | 26.6< | 3 | 5 |
| Typhloperipatus williamsoni | — | 1 | — | — | 3 | 1 | 1 | 3 | 21.1 | 1 | 4 |
| Peripatopsidae | | | | | | | | | | | |
| Metaperipatus blainvillei | — | 2 | — | — | 2 | 2 | 1 | 4 | 21.1 | 3 | 5 |
| Metaperipatus stesi | — | 2 | — | 20 | 3 | 3 | 1 | 3 | 21.1 | 1 | 2 |
| Opisthopatus cinctipes | — | 2 | — | — | 3 | 4 | 2 | 3 | 21.1 | 1 | 3 |
| Peripatopsis moseleyi | — | 1 | 24 | 6-10 | 3 | 4 | 2 | 3 | 21.1 | 1 | 3 |
| Peripatopsis capensis | 3.7-8 | 1 | — | 5-23 | 3 | 3 | 2 | 3 | 21.1 | 2 | 3 |
| Peripatopsis balfouri | — | 1 | — | — | 3 | 3 | 2 | 3 | 21.1 | 2 | 3 |
| Peripatopsis sedgwicki | — | 1 | — | 8-10 | 3 | 3 | 2 | 3 | 21.1 | 2 | 3 |
| Peripatoides novae-Zealandiae | — | 2 | — | — | 4 | 2 | 1 | 4 | 10 | 4 | 5 |
| Paraperipatus novae-britanniae | — | 2 | — | — | 5 | 1 | 1 | 2 | 26.6< | 5 | 5 |
| Ooperipatus sp. | — | 1 | — | 14 | 2 | 5 | 1 | 4 | 21.1 | 2 | 3 |
| Peripatoides orientalis | — | 2 | 36 | 53 | 2 | 3 | 1 | 4 | 21.1 | 2 | 3 |
| Euperipatoides leuckarti | — | 1 | — | — | 2 | 2 | 1 | 4 | 21.1 | 2 | 3 |
| Euperipatoides gifesi | 3.3 | — | — | — | 2 | 3 | 1 | 4 | 21.1 | 1 | 3 |

PI Parental investment as % of mothers body weight per young.
RF Reproductive frequency: I seasonal. 2 non-seasonal.
AF Age of female at first parturition. in months (lower values used for regression).
FEC Fecundity (number of young born each year per female, mean value used for regression).
QV Habitat vegetation type during the Quaternary, 18000 ybp. 1 seasonal forest. 2 cold deciduous arid forest. 3 cold deciduous forest. 4 cold deciduous moist forest. 5 seasonal tropical forest. 6 drought seasonal forest and grassland (aft Anonymous 1988).
VG Current habitat vegetation type: I tropical rain forest. 2 temperate forest. 3 Mediterranean. *4* tropical grassland. 5 temperate eucalyptus forest (after Anonymous 1979).
PH Photosymtheiic potential:I high. 2 medium after Anonymous 1988)
CL Climate type: I seasonal savanna. 2 hot damp. 3 warm with dry winters. *4* damp temperate (after Anonymous 1979).
TP Mean annual temperature, in Centigrades (after Anonymous (979).
MR Minimum rainfall per year: 1 :<2.54 cm. 2:2.54-5.08. 3:7.62-10.16. 4:10.16-20.32. 5:>20.32. (after Anonymous 1979).
RA Mean annual rainfall: I:<25.4 cm. 2:27.94-50.8. 3:53.34-101.6. 4:104.14-201 5:>203.2 (after Anonymous 1979).
— No data are available.

* Based on Manton 1938. Holiday 1944. Morrison 1946, Lavallard and Campiglia 1975, Morera *et al*. 1988. Campiglia and Lavallard. (973. (989. Havel *et al*. 1989, this paper and map sources indicated above.

**Maternal mass and reproductive characteristics:** An analysis of data in Lavallard and Campiglia (1975) indicates that a *P. acacioi* mother invests 7.9± 1.8(5-13) in each offspring ("Investment" = newborn's weight as % of mother's weight). Investment is predicted ($p<0.001$, $N=348$) by the equation I= (MW)(-0.0082)+ 13.78 where I=Investment and MW= maternal weight (Fig. 1).

The analysis also indicates that, in this species, fecundity is a function of maternal mass (Spearman c=0.50, $p<0.05$, N=l 17) predicted by the equation NY=0.0039MW÷0.866, where NY = number of young produced per parturition season and MW = maternal weight.

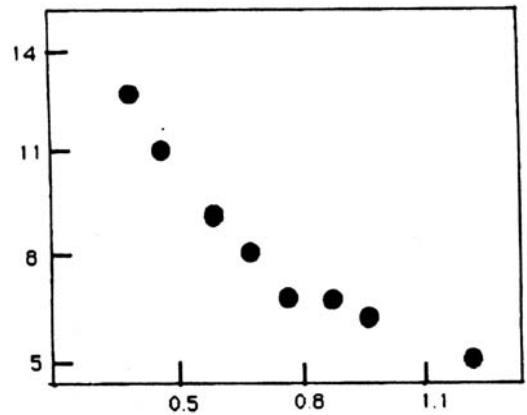

Fig. 1. *Peripatus acacioi:* correlation between parental investment (y axis as % of maternal weight represented by each offspring) and maternal weight in g (x axis). From data in Lavallard and Campiglia 1975.

TABLE 2

*Reponed sex ratios % onychophorans*

| Taxon | Females (%) | Source |
|---|---|---|
| Family Peripatidae | | |
| Undetermined neotropical species | 75 | Bouvier 1905 |
| *Peripatus juliformis* (Grenada) | 71 | Brues 1911* |
| *Oroperipatus eiseni* (Mexico) | 65 | Wheeler 1898* |
| *Macroperipatus torquatus* (Trinidad) | 65 | Read 1985 |
| *Macroperipatus trinidadensis:* nature (Trinidad) | 65 | Read 1985 |
| *Plicatoperipatus jamaicensis:* adults (Jamaica) | 62 | Havel *et al.* 1989 |
| *Peripatus acacioi* (Southern Brazil) | 61 | Monge-Nájera and Morera 1994 |
| *Epiperipatus biolleyi* (Costa Rica) | 54 | Lavallard and Campiglia 1973.1975 |
| *Macroperipatus torquatus:* (captive born) | 53 | Read 1985 |
| *Macroperipatus trinidadensis:* (captive born) | 53 | Read 1985 |
| *Plicatoperipatus jamaicensis:* embryos (Jamaica) | 50 | Havel et al. 1989 |
| *Peripatus heloisae* (Central Brazil) | 48 | Carvalho 1942 |
| Family Peripalopsidae | | |
| *Paropisthopatus umbrinus* (Chile) | 74 | Johow 1911* |
| *Peripatopsis capensis* (S. Africa) | 67 | Moseley 1874* |
| Undetermined Australian species | 67 | Steel 1896* |
| *Peripatopsis balfouri* (S. Africa) | 39 | Purcell 1900* |
| *Peripatopsis leonina* (S. Africa) | 31 | Purcell 1900* |

\* Source cited by Campiglia and Lavallard 1973 and Lavallard and Campiglia 1973. 1975

TABLE 3

*Basic morphometric data for several species of the phylum Onychophora. family Peripatidae*

| Species | Legs/females | | Length/females | | Legs/ males | | Length/males | |
|---|---|---|---|---|---|---|---|---|
| | Min. | Max. | Min. | Max. | Min. | Max. | Min. | Max. |
| *Oroperipatus equadorensis* | 39 | 39 | 69 | 69 | M | M | M | M |
| *Oroperipatus lankesteri* | 37 | 38 | 35 | 82 | 33 | 35 | 32 | 32 |
| *Oroperipatus tuberculatus* | 37 | 37 | 73 | 73 | 37 | 37 | M | M |
| *Oroperipatus quitensis* | 36 | 36 | M | M | 36 | 36 | 37 | 37 |
| *Oroperipatus cameranoi* | 34 | 36 | 40 | 55 | 32 | 32 | 34 | 34 |
| *Oroperipatus corradoi* | 26 | 29 | 14 | 60 | 24 | 27 | 14 | 25 |
| *Oroperipatus eiseni* | 27 | 28 | 30 | 57 | 23 | 26 | 20 | 23 |
| *Oroperipatus belli* | 28 | 28 | 43 | 43 | 25 | 25 | 15 | 15 |
| *Oroperipatus goudoti* | 27 | 28 | 27 | 27 | 24 | 24 | M | M |
| *Oroperipatus soratanus* | 32 | 32 | 41 | 41 | 28 | 28 | M | M |
| *Oroperipatus intermedius* | 32 | 32 | 37 | 37 | M | M | M | M |
| *Plicatoperipatus jamaicensis* | 43 | 43 | 25 | 65 | 35 | 35 | 25 | 25 |
| *Macroperipatus torquatus* | 41 | 42 | 100 | 150 | 41 | 42 | 100 | 100 |
| *Macroperipatus perrieri* | 32 | 32 | 51 | 51 | 28 | 28 | M | M |
| *Macroperipatus geayi* | 30 | 33 | 27 | 100 | M | M | 28 | 28 |
| *Macroperipatus ohausi* | 27 | 28 | 44 | 44 | M | M | M | M |
| *M. ohausi guianensis* | 27 | 27 | 58 | 58 | 24 | 24 | 33 | 23 |
| *Macroperipatus valerioi* | 34 | 34 | 85 | 115 | M | M | M | M |
| *Peripatus sedgwicki* | 29 | 32 | 25 | 60 | 28 | 30 | 23 | 30 |
| *Peripatus juliformi* | 33 | 34 | 36 | 75 | 29 | 30 | 14 | 16 |
| *P.juliformis swainsonae* | 31 | 34 | 16 | 65 | 28 | 30 | 23 | 17 |
| *P.juliformis danicus* | 31 | 33 | 26 | 45 | 26 | 28 | 9 | 21 |
| *Peripatus brölemanni* | 30 | 33 | 39 | 65 | 29 | 29 | 27 | 28 |
| *Peripatus dominicae* | 28 | 31 | 29 | 56 | 25 | 25 | 17 | 25 |
| *Peripatus antiguensi* | 31 | 31 | 38 | 44 | 24 | 26 | 18 | 21 |
| *P. dominicae juanensis* | 31 | 32 | 37 | 54 | 27 | 27 | 18 | 18 |
| *Peripatus heloisae* | 31 | 34 | M | M | 28 | 32 | M | M |
| *Epiperipatus brasiliensis* | 31 | 32 | 37 | 80 | 29 | 29 | 37 | 37 |
| *Epiperipatus imthurni* | 29 | 31 | 25 | 96 | M | M | M | M |
| *Epiperipatus evansi* | 28 | 28 | 32 | 58 | M | M | M | M |
| *Peripatus trinitatis* | 28 | 32 | 22 | 62 | 27 | 30 | 21 | 30 |
| *Epiperipatus edwardsii* | 29 | 34 | 23 | 56 | 28 | 29 | 25 | 30 |
| *Epiperipatus simoni* | 28 | 32 | 40 | 68 | M | M | M | M |
| *Oroperipatus balzani* | M | M | M | M | 26 | 27 | 27 | 31 |
| *Epiperipatus biolleyi* | 29 | 30 | 25 | 55 | 25 | 28 | 22 | 29 |
| *Epiperipatus nicaraguensis* | 32 | 32 | 46 | 46 | M | M | M | M |
| *Epiperipatus isthmicola* | 29 | 32 | 20 | 73 | 26 | 27 | 20 | 48 |
| *Epiperipatus hilkae* | 28 | 29 | 56 | 56 | 25 | 27 | M | M |
| *Mesoperipatus tholloni* | 24 | 27 | 52 | 60 | 23 | 24 | 38 | 38 |
| *Eoperipatus sumatranus* | M | M | M | M | 24 | 24 | 25 | 25 |
| *Eoperipatus weldoni* | 22 | 25 | 52 | 65 | 22 | 22 | M | M |
| *Eoperipatus horsti* | 24 | 25 | 46 | 46 | 23 | 24 | 40 | 40 |
| Family Peripatopsidae | | | | | | | | |
| *Ophistopatus cinctipes* | 16 | 16 | 7 | 50 | 16 | 16 | 6 | 36 |
| *Ophisthopatus roseus* | 18 | 18 | 30 | 42 | M | M | M | M |
| *Peripatus capensis* | 16 | 25 | 7 | 75 | 16 | 25 | 7 | 50 |
| *Peripatopsis alba* | 18 | 18 | 32 | 32 | M | M | 48 | 48 |
| *Peripatopsis balfouri* | 18 | 18 | 8 | 43 | 18 | 18 | 8 | 32 |
| *Peripatopsis capensis* | 18 | 18 | 9 | 70 | 18 | 18 | 6 | 54 |
| *Peripatopsis clavigera* | 17 | 18 | 7 | 52 | 17 | 18 | 7 | 36 |
| *Peripatopsis leonina* | 20 | 24 | 7 | 41 | 20 | 24 | 7 | 34 |
| *Peripatopsis moseleyi* | 21 | 25 | 11 | 75 | 20 | 24 | 9 | 50 |
| *Peripatopsis sedwicki* | 20 | 20 | 12 | 68 | 20 | 20 | 10 | 46 |
| *Metaperipatus blainvillei* | 20 | 22 | 9 | 65 | 19 | 22 | 9 | 40 |
| *Paropisthopatus costesi* | 16 | 16 | 20 | 36 | M | M | M | M |
| *Paropisthopatus umbrinus* | 16 | 16 | 20 | 70 | M | M | M | M |
| *Austroperipatus paradoxus* | 15 | 15 | 7 | 80 | 15 | 15 | 6 | 36 |
| *Euperipatoides leuckarti* | 15 | 15 | S | 40 | 15 | 15 | 4 | 29 |
| *Mantonipatus persiculus* | 15 | 15 | 8 | 33 | 15 | 15 | 8 | 20 |
| *Occiperipatoides gilesi* | 16 | 16 | 7 | 46 | 16 | 16 | 5 | 31 |
| *Ooperipatellus insignis* | 14 | 14 | 5 | 39 | 14 | 14 | 4 | 30 |
| *Ooperipatellus nanus* | M | M | 6 | 8 | M | M | 5 | 5 |
| *Oopuipalus oviparus* | 15 | 15 | 4 | 60 | 15 | 15 | 4 | 20 |

| Species | | | | | | | | |
|---|---|---|---|---|---|---|---|---|
| *Peripatoides indigo* | 15 | 15 | 8 | 76 | 15 | 15 | 44 | 44 |
| *Peripatoides novaezealandiae* | 15 | 15 | 12 | 58 | 15 | 15 | 10 | 39 |
| *Paraperipatus amboinensis* | 21 | 21 | 24 | 32 | M | M | M | M |
| *Paraperipatus ceramensis* | 21 | 22 | 13 | 55 | M | M | M | M |
| *Paraperipatus keiensis* | 24 | 25 | 27 | 48 | 22 | 23 | 25 | 33 |
| *Paraperipatus novaebritanniae* | 24 | 24 | 14 | 55 | 22 | 23 | 14 | 26 |
| *Paraperipatus papuensis* | 21 | 29 | 22 | 83 | 21 | 27 | 19 | 45 |
| *Paraperipatus schultzei* | 26 | 27 | 45 | 90 | 23 | 24 | 39 | 39 |
| *Paraperipatus stresemanni* | 23 | 25 | 25 | SO | M | M | M | M |
| *Paraperipatus vanheurni* | 25 | 27 | 15 | 60 | 21 | 21 | 24 | 24 |
| *Tasmanipatus barretti* | 15 | 15 | 23 | 40 | 15 | 15 | 36 | 36 |
| *Tasmanipatus anophthalmus* | 15 | 15 | 25 | 30 | 15 | 15 | 16 | 17 |

Key: Min= minimum, Max= maximum

M No data available. All lengths are in mm.
Sources: Bouvier 1905, 1907, Peek 1975, Ruhberg 1985, Read 1985, Morera and León 1986, Lavallard and Campiglia 1973, 1975, Ruhberg *et al.* 1991. Monge-Nájera and Morera 1994.

TABLE 4

Descriptive morphological statistics of Several onychophoran species (calculated from Table 3)

Peripatidae

| Variable | Cases | Mean | Stand. Dev .. | Minimum | Maximum |
|---|---|---|---|---|---|
| MALH | 38 | 63 | 21 | 27 | 150 |
| MALM | 27 | 31 | 16 | 15 | 100 |
| MAPH | 39 | 32 | 4 | 25 | 43 |
| MAPM | 32 | 29 | 4 | 22 | 42 |
| MILH | 38 | 40 | 19 | 14 | 100 |
| MILM | 27 | 27 | 17 | 9 | 100 |
| MIPH | 39 | 31 | 5 | 22 | 43 |
| MIPM | 32 | 28 | 5 | 22 | 41 |
| RALH | 38 | 23 | 22 | 00 | 73 |
| RALM | 27 | 3 | 27 | lO | 28 |
| RAPH | 39 | 2 | 2 | 00 | 5 |
| RAPM | 32 | 1 | 1 | 00 | 3 |

Peripatopsídae

| Variable | Cases | Mean | Stand. Dev .. | Minimum | Maximum |
|---|---|---|---|---|---|
| MALH | 32 | 53 | 19 | 8 | 90 |
| MALM | 26 | 35 | 12 | 5 | 54 |
| MAPH | 31 | 20 | 5 | 14 | 29 |
| MAPM | 24 | 19 | 4 | 14 | 27 |
| MILH | 32 | 15 | 10 | 4 | 45 |
| MILM | 26 | 15 | 13 | 4 | 48 |
| MIPH | 31 | 18 | 4 | 14 | 26 |
| MIPM | 24 | 18 | 3 | 14 | 23 |
| RALH | 32 | 38 | 21 | 00 | 73 |
| RALM | 26 | 20 | 15 | 00 | 48 |
| RAPH | 31 | 1 | 2 | 00 | 9 |
| RAPM | 24 | 1 | 2 | 00 | 9 |

Key:
MALH Maximum length, femates
MALM Maximum length, males
MAPH Maximum number of legs, females
MAPM Maximum number of legs, males
Mll..H Minimum lenght, females
MILM Mínimum len~ht, males
MIPH Minimum number of legs, females
MIPM Minimum number of legs. males
RALH Range of lenght, females
RALM Range of lenght, males
RAPH Range of number of legs, females
RAPM Range of number of legs, males

**Associations of length and number of legs:** Morphometric data appear in Table 3. There were few correlations in length values, the only case was that longer peripatopsid females were more variable in their length (Spearman c=0.92, n=32). When sexes are compared, females have higher values for minimum and maximum length and for length variability (Mann-Whitney U, p<0.01), with the exception of minimum length in peripatopsids (Mann-Whitney U, p>0.05).

Peripatopsid species with longer females also have longer males (Spearman c=0.70) but that was not the case with peripatids.

The minimum and maximum numbers of leg pairs were correlated (Spearman c≥0.79, n=24-39). Additionally, in peripatopsids, species with more legs are also more variable in the number of leg pairs (Spearman Test, females c=79, n=3 1; males c=84, n=24).

When comparing sexes, only peripatid females have higher values for minimum and maximum number of legs (Mann-Whitney U. p<0.01).

**Analysis of variation within populations:** The descriptive statistics for body length, weight and number of leg pairs, as well as descriptive regressions for some cases, appear in Table 6.

TABLE 5

*Morphometric data at the population level for 12 species of the family Peripatidae\**

I. Number of leg pairs

| Taxon | Sex | Mean | S.D. | Min. | Max. | N |
|---|---|---|---|---|---|---|
| *Epiperipatus biolleyi* | | | | | | |
| | Males | 27.76 | 0.60 | 27 | 30 | 58 |
| | Females | 30.33 | 0.66 | 28 | 32 | 95 |
| *Epiperipatus broadwayi* | | 31.38 | 1.40 | 29 | 34 | 21 |
| *Epiperipatus imthurni* | | | | | | |
| | Females | 30.35 | 0.56 | 29 | 32 | 119 |
| *Epiperipatus trinidadensis* | | 28.95 | 0.91 | 27 | 31 | 125 |
| *Macroperipatus torquatus* | | 40.15 | 1.2 | 37 | 42 | 652 |
| *Peripatus acacioi* | | | | | | |
| | Males | 25.60 | 0.59 | 24 | 27 | 348 |
| | Females | 27.70 | 0.62 | 26 | 30 | 450 |
| *Peripatus antiguensis* | | 28.73 | 2.05 | 24 | 31 | 11 |
| *Peripatus dominicae* | | 28.33 | 1.80 | 25 | 30 | 43 |
| *Peripatus eisenii* | | | | | | |
| | Males | 24.55 | 0.57 | 23 | 25 | 29 |
| *Peripatus heloisae* | | | | | | |
| | Males | 30.02 | 0.53 | 28 | 32 | 654 |
| | Females | 32.96 | 0.57 | 31 | 34 | 602 |
| *Peripatus juliformis* | | | | | | |
| | Males | 28.48! | 0.58 | 28 | 30 | 52 |
| | Females | 32.75 | 0.71 | 31 | 34 | 134 |
| *Plicatoperipatus jamaicensis* | | | | | | |
| | Males | 35.08 | 0.86 | 31 | 37 | 99 |
| | Females | 36.92 | 0.73 | 35 | 39 | 140 |

II. Weight (mg)

| | | | | | | |
|---|---|---|---|---|---|---|
| *Epiperipatus biolleyi* | Males | 1358 | 854.8 | 58 | 3369 | 94 |
| | Females | 2851 | 1905 | 100 | 8352 | 94 |
| *Peripatus acacioi* | | | | | | |
| | Males | 140.60 | 69.47 | 25 | 375 | 348 |
| | Females | 379.8 | 285.5 | 32 | 1350 | 450 |
| *Epiperipatus trinidadensis* | | 314.5 | 160.3 | 25 | 675 | 76 |

III. Length (mm)

| | | | | | | |
|---|---|---|---|---|---|---|
| *Epiperipatus biolleyi* | | | | | | |
| | Males | 38.49 | 8.62 | 18 | 55 | 53 |
| | Females | 52.27 | 13.25 | 18 | 75 | 84 |

*Key: SD= Standard Deviatiun. Min.= Minimum. Max.=maximum. N=sample size.

Sources for morphometric data: Campiglia and Lavallard 1973. Lavallard and Campiglia 1973. 1975. Read 1985. 1988. Havel *et al*. 1989. Monge-Nájera and Morera 1994.

The variabilities, calculated as (standard deviation/mean) x 100, were as follows:
Length *(E. biolleyi):* males 22.40, females 25.35 %.
Weight: *E. biolleyi* males 62.94, females 66.82: *P. acacioi:* males 49.41, females 75.17%.

Number of leg pairs: female variability ranges from 1.73 *(P. heloisae)* to 2.24 *(P. acacioi),* and in males the range is 1.76 *(P. heloisae)* to 2.45 % *(P. jamaicensis).* Higher variabilities *(e.g.* 7.13 %, *P. antiguensis)* were obtained in those species for which authors did not specify sex, suggesting that both sexes were included in their samples.

Species differ among themselves in number of legs and weight (Kruskal-Wallis Anova p=0.0000).

TABLE 6

*Descriptive regressions for length, weight and number of legs in two peripatid species*

*Eperipatus biolleyi*

Males (N=46)
legs= 27.4+0.0002 weight
Length= 26.2+0.0008 weight
Weight =-8 130+280 legs+50 length

Females (N=73)
Length= 129.6.3.2 legs+0.006 weight
Weight= - 0.00016+442 legs + 110 length

*Peripatus acacioi*

Males (N=348)
Legs= 25.3+0.0024 weight
Weigth= -732÷34 legs

Females (N=450)
Legs 28-0.0005 weight
Weight= 3476.7-I 12 legs

*E. biolleyi* regressions from Monge-Nájera and Morera (1994);
(hose of *P. acacioi* were calculated from data described in Table 5
and are in the same units.

**Within-species comparisons of females and males:** For all species with data for both sexes, these differ for each character (Mann-Whitney U, p=0.0000). Female *E. biolleyi* are 26 % longer than males. There are no equivalent data for the other species.

When comparing sexes, females are heavier: 47 % in *E. biolleyi* and 63 % in *P. acacioi.*

The percentages (%) by which females exceed males in number of legs are: *P. dominicae* 1.4, *P. jamaicensis* 5, *P. acacioi* 7.6, *E. biolleyi* 8.5 and *P. heloisae* 8.9.

DISCUSSION

**Family comparison:** Both families differed in important reproductive characteristics. It is not possible to distinguish, with current data, if these differences are of taxonomic origin or simply reflect the occurrence of peripatopsids in temperate habitats (compared with the tropical distribution of peripatids). A hypothesis for future evaluation is that peripatopsids are "ecologically derived" species (see Monge-Nájera l994b) which adapted to harsher environments by retarding reproduction, increasing fecundity and developing reproductive diapause. It is usually assumed that peripatids have a higher level of parental investment because of their complex reproductive physiology which includes viviparity (Morera *et al.* 1988, Monge-Nájera l994b). if the finding that mothers of both families invest similar proportions (by weight) in each offspring is not caused by the low sample size for peripatids, and if the higher fecundity of peripatopsids is taken into account, then these invest more annually than peripatids. However, this result may change if sample size is improved. Data about total investment in biomass per lifetime are needed for a strong conclusion.

It is normally accepted that peripatids are more variable in the number of leg pairs than peripatopsids (Campiglia and Lavallard 1973, Lavallard and Campiglia 1973). These results corroborate that belief for females and indicate that peripatids are less variable in length. However, there is no difference in the variability of number of legs between males of both families. A possible explanation is that high variability is typical of large animals. Peripatid females are often the largest onychophorans (despite the lack of significant statistical difference in mean values) and also the most variable thus, variability may reflect size rather than sex, family or degree of evolutionary change as was previously believed (see Campiglia and Lavallard 1973. Lavallard and Campiglia 1973). Even parthenogenetic *E. imthurni* have a range of four in the number of leg pairs (Read 1985).

**Character associations with habitat:** Reproductive seasonality, which is reflected in diapause, was previously explained with the rule "tropical non-seasonality, temperate seasonality" which had too many exceptions (see critique in Read 1985). This analysis shows seasonal diapause to be a function of mean annual rainfall, in accordance with the observation that humidity is more important in onychophoran ecology than temperature, light and vegetation (Ruhberg 1985, Monge-Nájera 1994a). Perhaps populations within a single species show geographical variation, with intermediate rain levels giving origin to year around reproduction with strong peaks, as in *P. acacioi* (Lavallard and Campiglia 1975).

Classical ecological theory predicts higher parental investment and more skewed sex ratios in harsher habitats (Steam 1976) but that relationship was not found with this sample and with the approach used.

**Maternal mass and reproductive characteristics:** This analysis of *P. acacioi* shows that larger mothers invest a lower percentage per young but nevertheless have larger individual offspring than small mothers (see Lavallard and Campiglia 1975, Read 1985, Havel *et al.* 1989). One hypothesis is that small females are forced to give birth to young of a fixed minimum size and as a result have reduced fecundity, a phenomenom apparently not noticed before. Possibly very small young are not viable.

**Associations of length and number of legs:** Length data are mostly based on alcohol preserved specimens, and sexual dimorphism could be overestimated because of differential contraction of males and females in alcohol (Monge-Nájera and Morera 1994). Thus, the reason why larger species are more variable in length and number of legs is unknown. However, the fact that larger species sometimes do not have the larger males may reflect specific differences in selection for small, rapid males and larger, more fecund females (Ghiselin 1974, Monge-Nájera and Morera 1994).

**Analysis of variation within populations:** In general, females are more variable than males in length and weight, but the difference is small in number of legs. Again, greater variabilities appear to be associated with large body size, as found for species morphometric ranges.

The number of legs is relatively constant, length is intermediate and weight is very variable. The number of legs is a genetically determined characteristic, not surprisingly less variable than length and weight, which depend on the activity and feeding condition of the animals, respectively (Campiglia and Lavallard 1973, Read 1985, Monge-Nájera and Morera 1994).

When sexes are compared within a species females can have 1.4-8.9% more leg pairs, be 26 % longer, and weigh 47-63 % more. This similar to other reports (Campiglia and Lavallard 1973, Lavallard and Campiglia 1973) and seems to reflect a selective trend toward small vagile males and large fecund females (Ghiselin 1974, Monge-Nájera 1994b, Monge-Nájera and Morera 1994).

The parameters considered here do not seem to have any correlation with latitude or other general geographical characteristics (according to maps by Anonymous 1979, 1988) of the sampled populations (Fig. 2; the names of collecting sites appear in the source literature).

Future workers should try to obtain similar data for the family Peripatopsidae, which is even more poorly known in this respect. The morphometric ranges of species differ significantly between families, and the same may be true of population parameters.

ACKNOWLEDGEMENTS

I thank Juan B. Chavarria, Maria I. González, Mario Blanco and Vernon Arias (University of Costa Rica, UCR) for their assistance, and especially Hilke Ruhberg (Hamburg University), Daniel Briceño, Bernal Morera Pedro León and Alvaro Wille (UCR) for suggestions to improve an earlier draft which became part of this paper. This study was financed by the author and UNED provided time to finish it.

RESUMEN

Un análisis cuantitativo de las características de algunos gusanos onicóforos mostró que en habitats con menores niveles de lluvia las hembras se reproducen a mayor edad, son más fecundas y tienden a presentar interrupción estacional ("diapausa") de la reproducción si la lluvia no supera los 200 cm anuales. Tales habitats corresponden a la familia Peripatopsidae. La proporción de sexos y La "inversión paterna" por neonato no se correlacionan con las condiciones ambientales generales. Las especies de cuerpo más grande a menudo son morfométricamente más variables, pero las especies con las hembras más largas no siempre tienen los machos más largos. Las hembras más grandes de *Peripatus acacioi* (Familia Peripatidae, (Brasil) producen más bebés y éstos son más pesados. También se hizo un estudio de variación poblacional. Para ello se analizó la literatura para aquellas especies de peripátidos de las cuales hay muestras de entre 11 y 798 individuos en cada una. En general, dentro de una población las hembras son más variables que los machos en longitud y peso, pero no tanto en el número de patas. La cantidad de patas es poco variable (1.73-2.45 %), la longitud es intermedia en variabilidad (22.4- 25.3%) y el peso es muy variable (49.41- 75.2). Cuando Se comparan los sexos dentro de una población, las hembras pueden tener 1.4-8.9 % más pares de patas, pesar 47-63 % más y ser 26 % más largas.

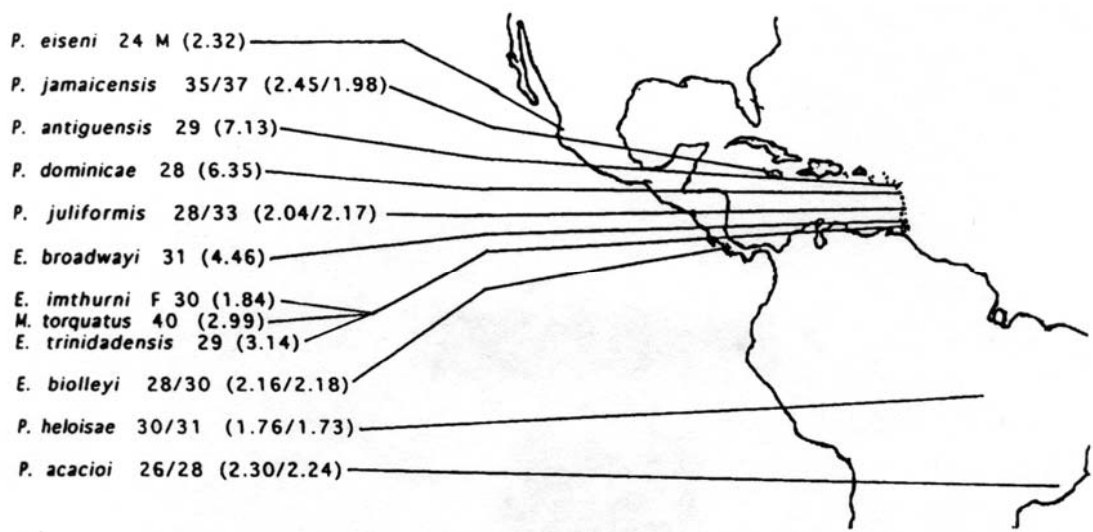

**Fig.** 2. Geographic distribution in the Neotropics of population means for number of leg pairs in the onychophoran family Pcripatidae. according to Table 5. Left: males, right: females. When source did no indicate sex, a single number appears; F females. M males. In parenthesis: variability measured as: (Standard deviation/Mean) a 100; same format as means.